\documentclass[aps,prl,twocolumn,showkeys,superscriptaddress]{revtex4-2}

\usepackage{placeins}                   
\usepackage[artemisia]{textgreek}       
\usepackage{amssymb}       
\usepackage{gensymb}
\usepackage{amsmath} 				    
\usepackage{latexsym}                   
\usepackage{nicefrac}                   

\usepackage{graphicx}                   
\usepackage{epsfig}                     
\usepackage{epstopdf}                   
\usepackage{dcolumn}                    

\usepackage{lipsum}                     
\usepackage{color}                      

\usepackage{subfigure}                  
\usepackage{blindtext}                  
\usepackage[ulem=normalem]{changes}     
\usepackage{float}                      
\usepackage{array}                      
\usepackage{soul}                       
\usepackage[T1]{fontenc}                
\usepackage[hidelinks]{hyperref}        

\usepackage[separate-uncertainty=true,multi-part-units=single]{siunitx}

\DeclareSIUnit\sq{\ensuremath{\Box}}
\DeclareSIUnit\bar{bar}
\DeclareSIUnit\angstrom{\text{Å}}
\DeclareSIUnit{\atpercent}{at\%}

\newcolumntype{P}[1]{>{\centering\arraybackslash}p{#1}}
\hypersetup{
    colorlinks,
    linkcolor={red!50!black},
    citecolor={blue!50!black},
    urlcolor={blue!80!black}
}

\newcolumntype{L}[1]{>{\raggedright\arraybackslash}p{#1}}
\newcolumntype{C}[1]{>{\centering\arraybackslash}p{#1}}
\newcolumntype{R}[1]{>{\raggedleft\arraybackslash}p{#1}}

\begin{document}

\title{Development of a Nb-based semiconductor-superconductor hybrid 2DEG platform}

\author{Sjoerd Telkamp}
\affiliation{Solid State Physics Laboratory, ETH Z\"urich, CH-8093 Z\"urich, Switzerland}
\affiliation{Quantum Center, ETH Z\"urich, 8093 Z\"urich, Switzerland}
\author{Tommaso Antonelli}
\affiliation{Solid State Physics Laboratory, ETH Z\"urich, CH-8093 Z\"urich, Switzerland}
\affiliation{Quantum Center, ETH Z\"urich, 8093 Z\"urich, Switzerland}
\affiliation{IBM Research Europe - Z\"urich, 8803 Rüschlikon, Switzerland}
\author{Clemens Todt}
\affiliation{Solid State Physics Laboratory, ETH Z\"urich, CH-8093 Z\"urich, Switzerland}
\affiliation{Quantum Center, ETH Z\"urich, 8093 Z\"urich, Switzerland}
\author{Manuel Hinderling}
\affiliation{IBM Research Europe - Z\"urich, 8803 Rüschlikon, Switzerland}
\author{Marco Coraiola}
\affiliation{IBM Research Europe - Z\"urich, 8803 Rüschlikon, Switzerland}
\author{Daniel Haxell}
\affiliation{IBM Research Europe - Z\"urich, 8803 Rüschlikon, Switzerland}
\author{Sofieke C. ten Kate}
\affiliation{IBM Research Europe - Z\"urich, 8803 Rüschlikon, Switzerland}
\author{Deividas Sabonis}
\affiliation{IBM Research Europe - Z\"urich, 8803 Rüschlikon, Switzerland}
\author{Peng Zeng}
\affiliation{Scientific Center for Optical and Electron Microscopy (ScopeM), ETH Zürich, CH 8093 Zürich, Switzerland.}
\author{R\"udiger Schott}
\affiliation{Solid State Physics Laboratory, ETH Z\"urich, CH-8093 Z\"urich, Switzerland}
\author{Erik Cheah}
\affiliation{Solid State Physics Laboratory, ETH Z\"urich, CH-8093 Z\"urich, Switzerland}
\author{Christian Reichl}
\affiliation{Solid State Physics Laboratory, ETH Z\"urich, CH-8093 Z\"urich, Switzerland}
\affiliation{Quantum Center, ETH Z\"urich, 8093 Z\"urich, Switzerland}
\author{Fabrizio Nichele}
\affiliation{IBM Research Europe - Z\"urich, 8803 Rüschlikon, Switzerland}
\author{Filip Krizek}
\affiliation{Institute of Physics, Czech Academy of Sciences, 162 00 Prague, Czech Republic}
\author{Werner Wegscheider}
\affiliation{Solid State Physics Laboratory, ETH Z\"urich, CH-8093 Z\"urich, Switzerland}
\affiliation{Quantum Center, ETH Z\"urich, 8093 Z\"urich, Switzerland}
\begin{abstract}
    Semiconductor-superconductor hybrid materials are used as a platform to realise Andreev bound states, which hold great promise for quantum applications. These states require transparent interfaces between the semiconductor and superconductor, which are typically realised by in-situ deposition of an Al superconducting layer. Here we present a hybrid material based on an InAs two-dimensional electron gas (2DEG) combined with in-situ deposited Nb and NbTi superconductors, which offer a larger operating range in temperature and magnetic field due to their larger superconducting gap. We overcome the inherent difficulty associated with the formation of an amorphous interface between III-V semiconductors and Nb-based superconductors by introducing a 7 nm Al interlayer. The Al interlayer provides an epitaxial connection between an in-situ magnetron sputtered Nb or NbTi thin film and a shallow InAs 2DEG. This metal-to-metal epitaxy is achieved by optimization of the material stack and results in an induced superconducting gap of approximately 1 meV, determined from transport measurements of superconductor-semiconductor Josephson junctions. This induced gap is approximately five times larger than the values reported for Al-based hybrid materials and indicates the formation of highly-transparent interfaces that are required in high-quality hybrid material platforms.
\end{abstract}
\maketitle
\section{Introduction}

%

Semiconductor-superconductor hybrid materials are used as a platform to realise sub-gap states, which are quasi-particles that arise when a low dimensional semiconductor is confined between two superconducting leads. The physics of these bound states, such as trivial Andreev bound states or more elusive topologically protected zero-energy modes, has been recently investigated in various semiconductor-superconductor hybrid systems ranging from quasi 1-D nanowires to 2-D electron systems \cite{Lutchyn2018,PhysRevB.52.1995,Shabani2016,VanWoerkom2017,Pillet2010}. Examples of experimental realizations using these hybrid materials include Andreev spin qubits  \cite{Zazunov2003,Hays2021,Pita-Vidal2023}, multi-terminal devices \cite{Coraiola2023,banszerus2024,banszerus2024_1,coraiola2024}, superconducting diodes \cite{Baumgartner2022,banszerus2024} and investigations of Majorana bound states \cite{vanmourik2012,Albrecht2016,TenHaaf2024}. A driving force behind this research field is the promise of exploiting these states for quantum computing applications. 
\begin{figure*}
\centering
\setlength\fboxsep{0pt}
\setlength\fboxrule{0pt}
\fbox{\includegraphics[width=6.5in]{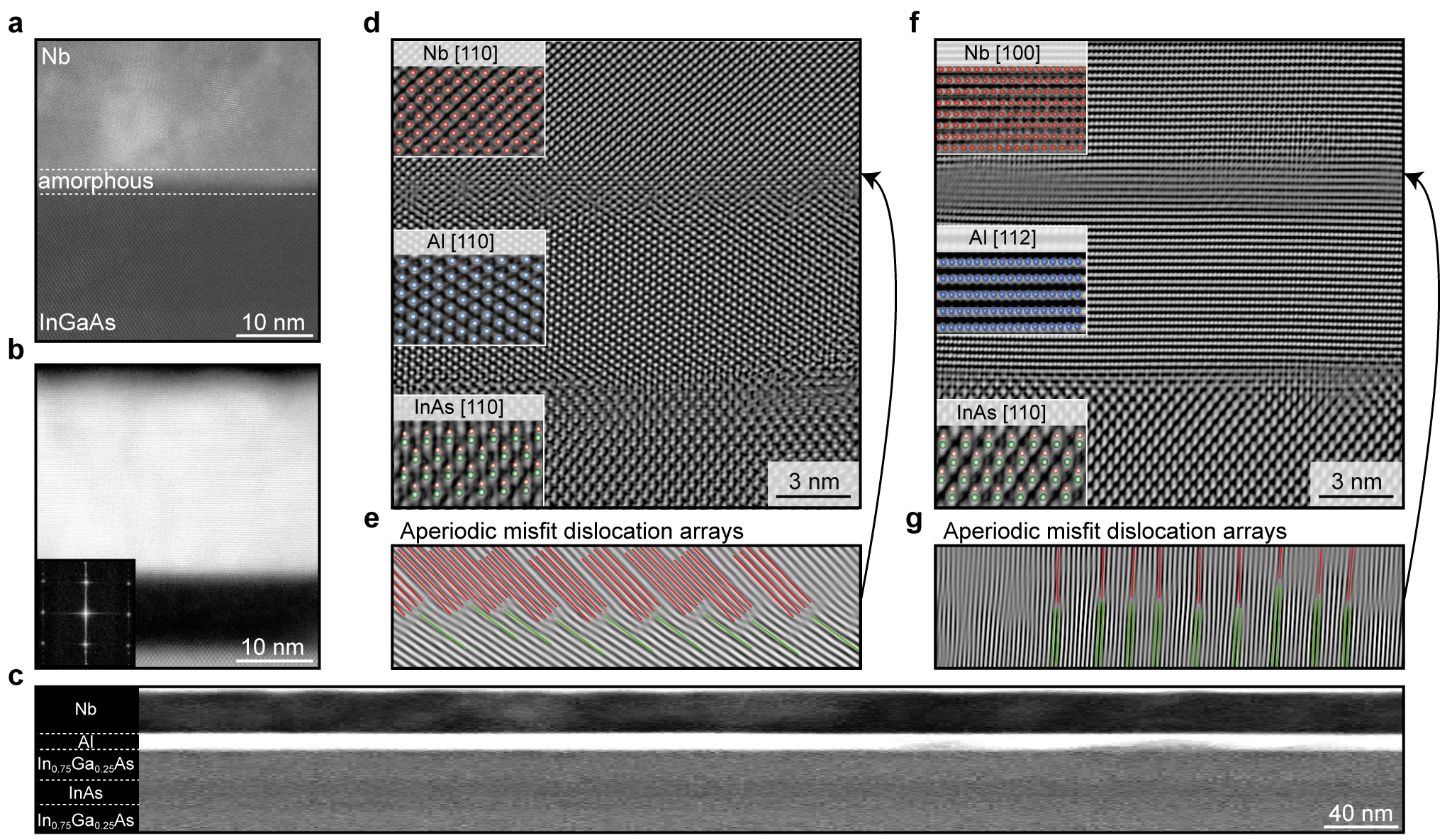}}
\caption{\textbf{STEM images showing the development of epitaxial interfaces between the semiconductor heterostructure and the Al and Nb superconducting layers.} \textbf{a} HAADF STEM image of the interface in case of Nb deposition directly the semiconductor surface. An amorphous layer at the interface is indicated by dashed lines. \textbf{b} Angular Dark Field STEM image showing InAs heterostructure, the Al and the Nb superconducting layer together with the fast Fourier transform (FFT) of the Nb, which highlights its single crystalline structure. \textbf{c} Overview STEM image of the upper part of the material stack, showing a homogeneous and defect free Al interlayer. \textbf{d} High-resultion HAADF STEM images showing the details of the semiconductor/Al and the Al/Nb interfaces, for a grain where the [110] projection of the Nb matches the [110] projection of the substrate. To identify the observed grain orientations the obtained STEM data is overlaid with crystallographic models. \textbf{e} Bragg-filtered portion of the interface in \textbf{d}, highlighting the presence of misfit dislocations that occur at the Nb/Al interface.  \textbf{f} Details of the semiconductor/Al and the Al/Nb interfaces, for a grain where the [100] projection of the Nb matches the [110] projection of the substrate.  \textbf{g} Bragg-filtered portion of the interface in \textbf{f}, highlighting the presence of misfit dislocations that occur at the Nb/Al interface.}
\label{f1}
\end{figure*}

Most experiments rely on the in-situ deposition of epitaxial Al films on top of high spin-orbit semiconductors such as InAs or InSb \cite{Chang2015,Moehle2021}. Recently, promising results were shown using other in-situ deposited superconductors, such as Sn \cite{Pendharkar2021}, Pb \cite{Kanne2021}, V \cite{Bjergfelt_2019} or Nb \cite{Perla2021}, which offer a larger operating range in temperature and magnetic field. This allows the possibility to investigate the coupling between superconductivity and quantum states in the two-dimensional electron gas (2DEG) available only at high magnetic fields and has the potential to improve the overall performance of semiconductor-superconductor hybrid materials. A high magnetic field regime that is of particular interest is the parameter space in which the quantum Hall effect (QHE) and superconductivity coexist \cite{Amet2016,Hatefipour2022,Guiducci2019}.

A crucial aspect of these semiconductor-superconductor hybrid structures is the quality of the interfaces \cite{Frolov}. This entails a highly ordered and electronically transparent interface without any chemical intermixing or impurities. High interface quality combined with physical proximity enable a large superconducting gap induced in the semiconductor that is free of sub-gap states. It was realized in 2015 by Chang et. al. \cite{Chang2015} that the most effective approach to obtain such high-quality interfaces is by in-situ deposition of the superconductor. In contrast to widely available ex-situ deposition methods, this prevents oxidation of the semiconductor and does not require any surface cleaning process that reduces the interface transparency. 

The most commonly used superconductor, Al, is compatible with thermal evaporation approaches such as molecular beam epitaxy (MBE) and has the possibility to form an epitaxial interface to III/V semiconductors such as InSb and InAs \cite{Thomas2019,Shabani2016}. Superconductors with a larger superconducting gap, such as Nb and NbTi, are less suitable for in-situ deposition due to their lower vapor pressure even at high temperature. Therefore, they typically require other deposition techniques such as electron beam evaporation or magnetron sputtering, which can be difficult to combine with the high-purity III-V semiconductor MBE systems used to grow hybrid 2-D materials \cite{Shabani2016,Cheah2023}. Additionally, amorphous interfaces due to chemical intermixing have been frequently reported when depositing superconductors such as Nb directly on the semiconductor, even if the deposition is done in-situ \cite{Todt2023,Perla2021,Gusken2017}. 

In this work, the incompatibility of III-V semiconductors and Nb-based superconductors is overcome by a novel approach of introducing a thin Al interlayer. This method results in the formation of a high-quality interface between the semiconducting and metallic parts of the structure. Subsequent deposition of Nb (or NbTi) on the Al interlayer improves the superconducting properties of the material stack and enforces a direct epitaxial relation between Al and Nb. We present a detailed scanning transmission electron microscopy (STEM) study of the different interfaces as well as optimization of the material stack. To demonstrate the suitability of the developed material platform for hybrid quantum devices, the transport behavior of Josephson Junctions (JJs) is analysed and compared to standard InAs/Al materials. Additionally, the uncommon formation of a high-quality epitaxial interface between two superconductors in our material stack gives an interesting perspective for future studies of the reverse proximity effect or superconducting gap engineering \cite{Mcwen2024,Diamond2022,Marchegiani2022}.


\section{In-situ superconductor deposition and metal-to-metal epitaxy}
\label{epitaxy}
The InAs semiconductor heterostructure is grown by MBE on a InP wafer. A step-graded buffer structure is grown to accommodate the lattice mismatch between InAs and InP and allow for dislocations to occur below the quantum well region. The 2DEG region consists of a bottom barrier, quantum well (QW) and a 13 nm InGaAs top barrier. The Al layer is grown in the same MBE system after cooling the sample to approximately $-20$  $\degree $C. The Nb top layer is subsequently deposited in-situ by magnetron sputtering in a custom-made metal deposition chamber with a base pressure lower than $2\times 10^{-10}$ mbar without breaking the ultra-high vacuum (UHV). The Nb and NbTi films are deposited at room temperature, with a rate of 2.6 \AA/s and at a pressure of $8.8\times 10^{-3}$ mbar. More details on the MBE growth process and metal deposition are reported in our previous work \cite{Cheah2023,Todt2023}. 

Figure \ref{f1}(a) shows a high-angle-annular dark field (HAADF) STEM image of the interface between Nb and the InGaAs top barrier when the Nb is deposited directly on the semiconductor heterostructure. An amorphous interlayer is formed at the interface which is typical for direct deposition of Nb on a semiconductor surface containing As and attributed to the formation of a NbAs compound \cite{Todt2023,Perla2021,Gusken2017}. This degree of chemical intermixing is known to be detrimental for the electronic interface transparency and will inhibit any epitaxial relationship. 

We avoid the formation of the amorphous interface by using epitaxial Al as an interlayer between the Nb and the 2DEG. This epitaxial Al interlayer can be penetrated by the Nb superconducting wave function due to the proximity effect enabled by high-quality interfaces. A thin Al film is a natural choice due its relatively low lattice mismatch to the semiconductor and already developed deposition procedure \cite{Cheah2023}. Figures \ref{f1}(b, c) show an overview of the InGaAs, Al and Nb layers on different length scales to evidence the formation of clean interfaces between the materials. The HR STEM images in Fig. \ref{f1}(d, f) show the epitaxial relationship between the materials. It is apparent that we have not only obtained an epitaxial interface between the InAs and the Al (which was already established \cite{Shabani2016,Cheah2023}), but that there is also a clear epitaxial relationship between the Nb and Al. This metal-to-metal epitaxy between the Al and the in-situ deposited Nb is, to the best of our knowledge, not yet reported in literature.

In the epitaxial Al film we observe the formation of two different grains with two different crystallographic orientations ([110] and [112]). As shown in Figs. \ref{f1}(d, f) these two different Al grains force the Nb to adapt two distinct crystallographic orientations ([110] and [100]). The mismatch in lattice constants between Nb and Al is compensated by the formation of aperiodic misfit dislocations at the interface, which are indicated in Figs. \ref{f1}(e, g). We observe the spacing of these dislocations to be on the order of 1 nm. The identification of two different grain combinations of Nb ([110] and [100]) and Al ([110] and [112]) is further substantiated by the XRD data in appendix Fig. \ref{xrd}. 

\section{Optimization of the Al interlayer}
\begin{figure}
 
    \centering
    \includegraphics[width = 1\linewidth]{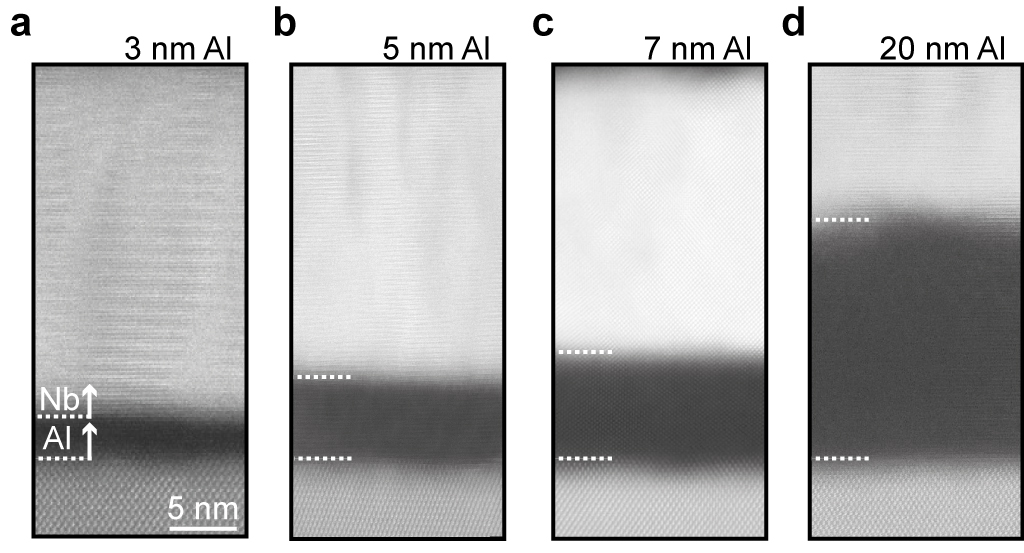}
    
    \caption{\textbf{The effect of Al thickness on the material interfaces}. \textbf{a} HAADF STEM image of the interfaces for a 3 nm thick Al layer that is not completely crystalline. \textbf{b}, \textbf{c} and \textbf{d} show the exact same material stack but with an Al thickness of 5 nm, 7 nm and 20 nm respectively.}
    \label{Alseries}
\end{figure}

One of the key parameters of this material stack is the thickness of the Al layer. The Al should be as thin as possible in order for the superconducting properties of the Nb to dominate. However, the thickness of the Al layer could also have an influence on the crystal properties of the material stack. Figure \ref{Alseries} shows STEM images from a series of four wafers with different Al thickness. The interface in Fig. \ref{Alseries}(a) indicates that if the Al layer is thinner than 3 nm it no longer has a crystalline structure and becomes amorphous, which can be expected to have a detrimental effect on the interface transparency. A possible explanation for this could be degradation of the thin Al film during UHV transfer. The interfaces shown in Fig. \ref{Alseries}(d) suggest that too thick Al leads to interface degradation, possibly related to an increase in strain, in addition to having less favourable superconducting properties. Based on this growth series, we established the optimum Al thickness to be between 5 and 7 nm. 

Another important feature of the material stack is the capping layer that prevents In migration through the Al layer. The In can diffuse from the semiconductor top barrier region and forms triangular zinc-blende inclusions in the Al layer (see Appendix \ref{A2:indium}). Partially-localized recrystallization (most likely into InAlAs) of the Al film and underlying InGaAs negatively affects both the superconducting properties and the confinement of the 2DEG via local modifications of the potential landscape.  The In diffusion can be suppressed by capping the InGaAs with a few monolayers (MLs) of GaAs. However, this capping layer affects the electronic properties of the semiconductor \cite{pauka2020,hatke2017} and the grain structure and lattice constant of the Al film \cite{Cheah2023,elbaroudy2024}. We therefore investigate the effect of the capping layer on the structure of the material stack. 

\begin{figure}
 
    \centering
    \includegraphics[width = 1\linewidth]{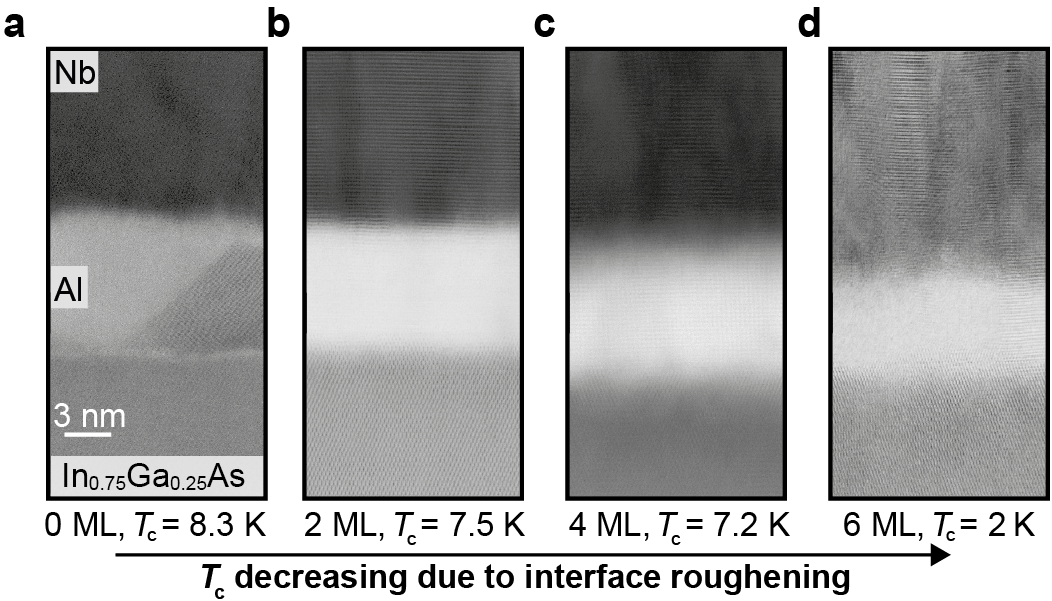}
    
    \caption{\textbf{GaAs capping of the semiconductor structure.}  \textbf{a} BF STEM images of the material stack with 7 nm Al and without GaAs capping. The dark triangular region within the Al layer corresponds to In migration into the Al interlayer and formation of a triangular inclusion. \textbf{b} Identical structures with 2 monolayers, \textbf{c} 4 monolayers and \textbf{d} 6 monolayers of GaAs deposited between the semiconductor and Al film. The addition of GaAs capping layers prevents In migration, but has a roughening effect on both the InAs/Al and Al/Nb interfaces.}
    \label{GaAs}
\end{figure}

Figure \ref{GaAs} shows a series of STEM images demonstrating the effect of the thickness of the capping layers on the obtained interfaces. The series consists of 4 growth runs in which the thickness of GaAs capping layers was increased from 0 to 6 monolayers while keeping all other growth parameters the same. The STEM image in Fig. \ref{GaAs}(a) confirms the aforementioned effect of indium diffusion in case of a direct interface between InGaAs and Al. For more than 4 MLs of GaAs both the InGaAs-Al and consequently Al-Nb interfaces deteriorate, most likely due to large amount of compressive strain introduced by GaAs as it exceeds the critical thickness. This is reflected by a rapid decrease of transition temperature ($T_{\rm c}$) with the number of deposited GaAs MLs. The reduction in $T_{\rm c}$ is likely related to oxidation along an increased density of crystallographic boundaries of Nb \cite{halbritter2005transport}. We find an optimum of 2 GaAs MLs, which suppresses the In diffusion while the $T_{\rm c}$ is only minimally reduced to 7.5 K. 

\section{Expanding the approach to N$\text{b}$T$\text{i}$}
\begin{figure}
 
    \centering
    \includegraphics[width = 1\linewidth]{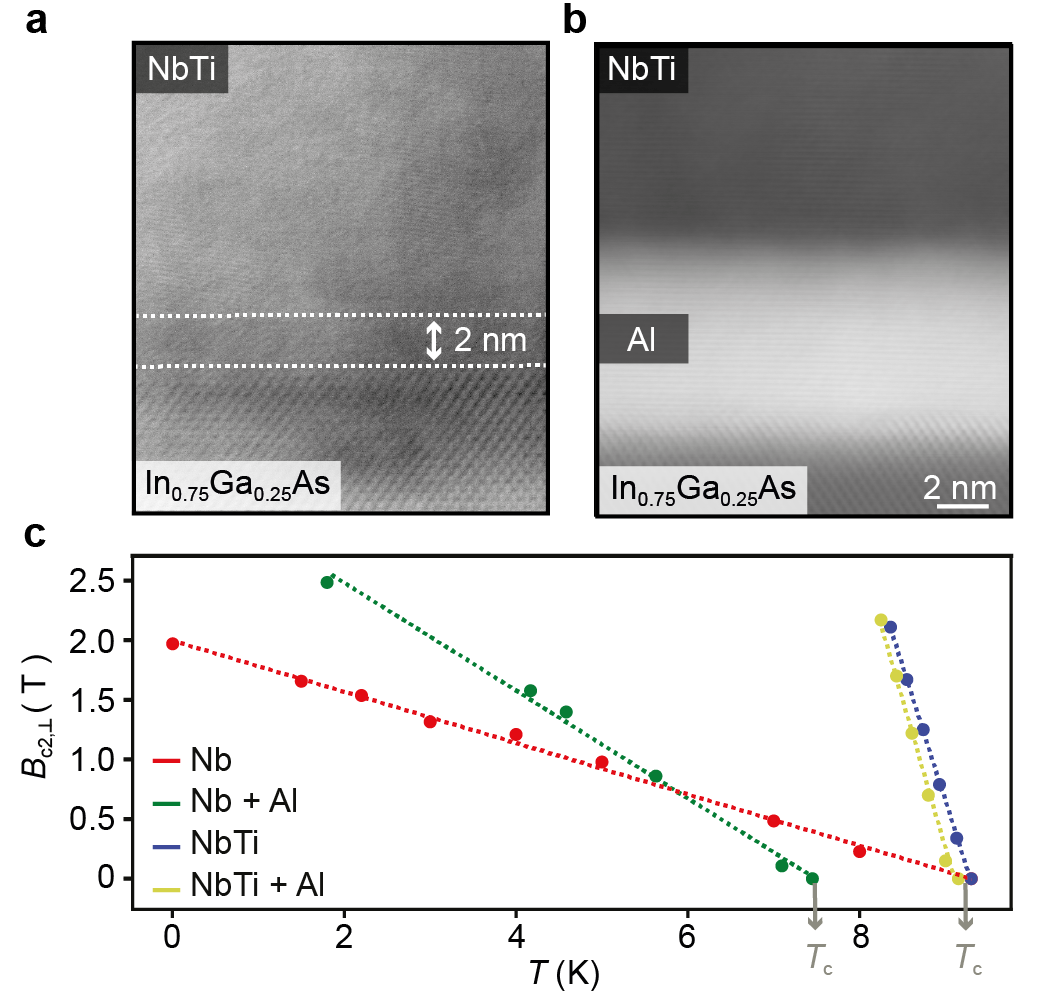}
    
    \caption{\textbf{Material stack with NbTi as the top superconductor}. \textbf{a} BF STEM images showing the interface of NbTi deposited directly on the semiconductor heterostructure, where a fully amorphous region is highlighted by the dashed lines. \textbf{b} BF STEM images of the high-quality interfaces formed due to insertion of the epitaxial Al interlayer. \textbf{c} Critical magnetic field as a function of temperature for Nb, Nb with Al, NbTi and NbTi with Al obtained from DC transport measurements. The substrate for all four films is the optimised semiconductor heterostructure with the InAs 2DEG and 2ML of GaAs capping. The grey arrows indicate the critical transition temperatures.}
    \label{NbTi}
\end{figure}

The presented approach of depositing a high-$B_{\rm c}$ superconductor on top of the epitaxially grown Al can be extended to other superconductors. In Fig. \ref{NbTi} we investigate the material stack in case of NbTi as the top superconductor. Similar to Nb, NbTi forms an amorphous interlayer when directly deposited on the semiconductor heterostructure, as is shown in Fig. \ref{NbTi}(a). However, high-quality interfaces are obtained if we insert a 7 nm Al interlayer. Figure \ref{NbTi}(b) shows that there are no signs of chemical intermixing and a similar epitaxial relationship between the NbTi and the Al is observed with the dominant grain orientation comparable to Fig. \ref{f1}(f). This is likely due to the similarity in the lattice constant of NbTi compared to Nb, with a relative difference of less than $< 0.5 \%$ \cite{Gajda2019}.

The out-of-plane critical magnetic field as a function of temperature of a 60 nm NbTi film, with and without Al 7 nm Al, are shown in Fig. \ref{NbTi}(c) and compared to Nb. All four films are deposited on the optimized semiconductor heterostructure. All of these films far exceed the known critical temperature and out-of-plane magnetic field of state-of-the-art epitaxial Al films of 1.5 K and 164 mT respectively \cite{Mayer2019}. The Nb films with Al show a slightly reduced $T_{\rm c}$ of 7.5 K while the $B_{\rm c}$ at low temperature is increased compared to Nb without Al. Furthermore, both NbTi films have a much larger critical magnetic field compared to the Nb films.  The $T_{\rm c}$ of the NbTi is not significantly affected by the presence of the Al layer.   

We attribute the increase in $B_{\rm c}$ for the Nb with Al films at low temperature to vortex pinning at the Nb-Al interface. The decrease in $T_{\rm c}$ of the Al-Nb metal stack compared to the Nb films is explained by the inverse proximity effect. This effect can be thought of as part of the Cooper pair wave function leaking into the Al and has been observed in other experiments, for example, in Nb-Ni bilayers \cite{Skryabina2019} but also Ti-Au bilayers \cite{Nagayoshi2022}. It can reduce the $T_{\rm c}$ when the film thickness is similar or smaller than the coherence length of the superconductor. Consequently, it does not affect the $T_{\rm c}$ of the NbTi due to its shorter coherence length of around 6 nm \cite{flukiger2012} compared to 38 nm for Nb \cite{PhysRevB.50.6307}.

\section{Transport measurements on a Josephson Junction}

The induced superconducting gap is estimated via multiple Andreev reflection (MAR) measurements in planar JJs. Devices are fabricated from the optimized heterostructure with 2 MLs GaAs, 7 nm Al and 60 nm of NbTi. The choice of NbTi is motivated by its larger $B_{\rm c}$ and $T_{\rm c}$, as well as the fact that NbTi proved more resilient against oxidation during fabrication. More details on the device fabrication process are provided in the Appendix \ref{JJfab}. 

\begin{figure*}
\centering
\setlength\fboxsep{0pt}
\setlength\fboxrule{0pt}
\fbox{\includegraphics[width=6.8in]{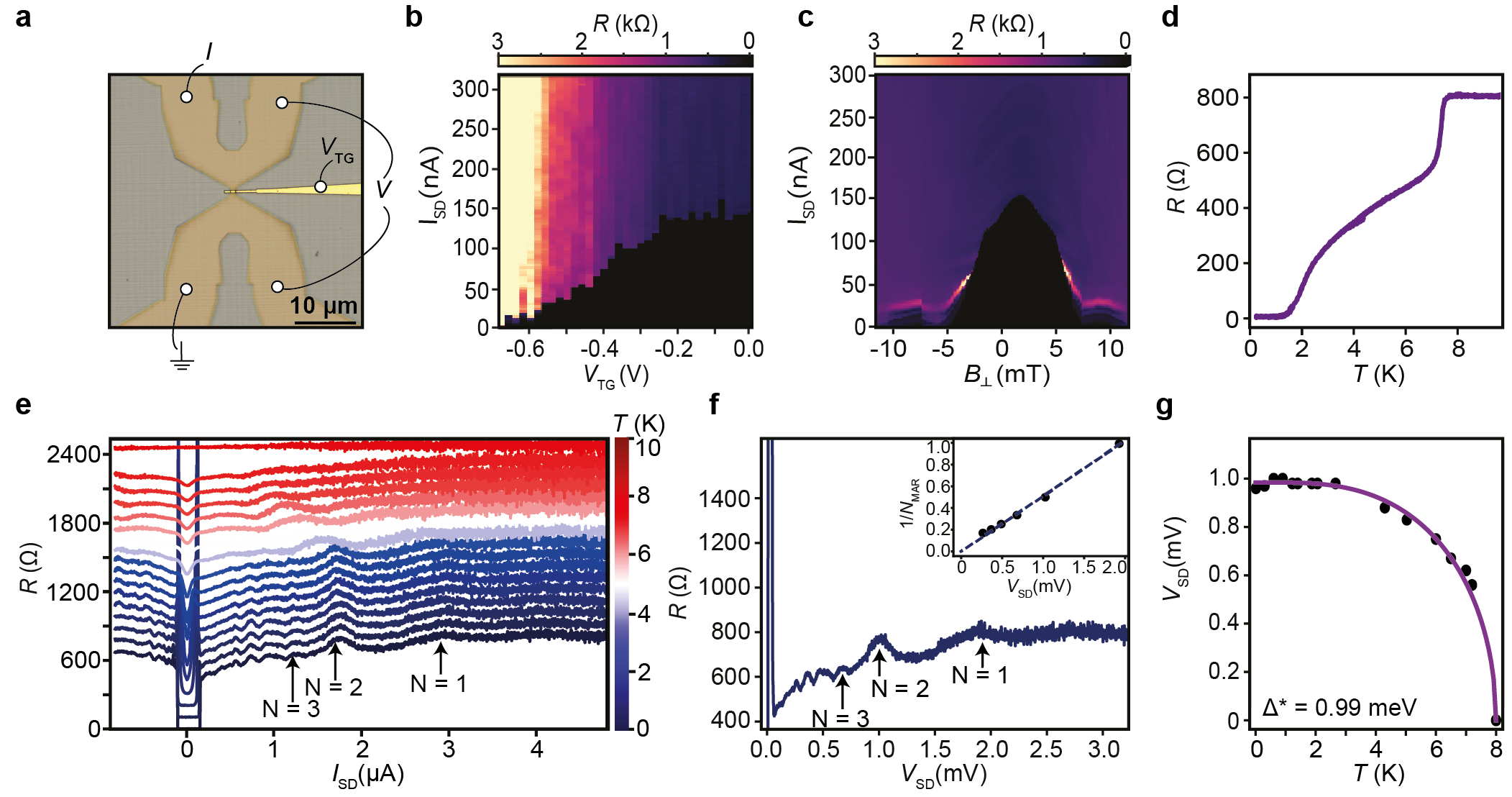}}
\caption{\textbf{Transport measurements of a Josephson Junction fabricated from this material.} \textbf{a} An optical image of the characterized device, where the source-drain electrodes, top-gate and the AC voltage probes are labeled. \textbf{b} The dependence of the resistance on the bias current and gate voltage. \textbf{c} The dependence of the resistance $R$ on the bias current and applied magnetic field. \textbf{d} The resistance of the JJ as a function of temperature at zero DC bias current. In \textbf{e} the resistance of the JJ as a function of DC bias voltage for increasing temperature is plotted. Indicated are the peaks that we assign to MAR. Traces are offset for clarity. Figure \textbf{f} shows the resistance at 600 mK as a function of $V_{\rm SD}$ with in the figure inset the first five MAR peaks plotted as a function of 1/$N_{\rm MAR}$. The peak locations are fitted with an induced gap of 0.99 meV. In \textbf{g} the temperature related shift of the $N=2$ MAR peak is plotted and fitted with equation \ref{BCSstar}. }
\label{f6}
\end{figure*}

The measurement configuration is illustrated on the optical micrograph of the device in Fig. \ref{f6}(a). The JJ is designed to be 100 nm in length and 1 $\mu$m in width and has four leads that are used to send a current through the junction and measure the voltage drop using standard lockin amplifier techniques. A top gate electrode covers the area of the junction (separated by a dielectric of 3 nm Al$_2$O$_3$ and 15 nm HfO$_2$) and allows for control of charge carrier density in the 2DEG. An AC excitation current of 4 nA at a frequency of 233 Hz is used which flows trough the junction. All measurements are carried out in a dilution refrigerator with a base temperature of below 10 mK. 

Figure \ref{f6}(b) shows the differential resistance $R=I_{\rm AC}/V_{\rm AC}$ of the JJ as a function of top gate voltage ($V_{\rm TG}$) and DC bias current ($I_{\rm SD}$). The supercurrent through the device is tuned by electrostatic gating and quenched for $V_{\rm g}= -0.7$ V. This indicates the supercurrent indeed flows through the 2DEG and can be pinched off by depleting the quantum well. In Fig. \ref{f6}(c), $R$ is shown as a function of bias current and out-of-plane magnetic field. The transition between the superconducting state and normal state in the device shows a Fraunhofer interference pattern. 
The observed Fraunhofer pattern is qualitatively similar to a planar Al based JJ of similar dimensions \cite{Haxell2022}. 
Based only on the device geometry, the first order minima should be offset from first order maximum of interference by 20 mT. We observe a smaller shift of 5 mT, which indicates a significant contribution of the flux focusing effect \cite{Suominen2017}. Additionally, the maximum in the supercurrent is found at 1.5 mT due to stray fields in the measurement setup. For all other measurements, we apply a field of 1.5 mT to compensate for these stray fields in the setup.  

The temperature dependence of $R$ at $I_{\rm SD} = 0$ is shown in Fig. \ref{f6}(d). The differential resistance decreases in a two-step manner as a function of temperature: there is a sharp decrease in $R$ at 7.5 K, followed by a gradual decrease in $R$ until a second sharp decrease at 2.5 K. Zero resistance is observed for $T<1.2$ K. The step at 7.5 K corresponds to the critical temperature of NbTi, indicating that the device is affected by the presence of NbTi in addition to the Al interlayers. Our results show good agreement with previous results on similar structures \cite{Drachmann2017a}. 

Figure \ref{f6}(e) shows how $R$ depends on the bias current for various temperatures. These traces are characterized by a dip in $R$ around zero current bias that reaches zero for temperature smaller than 1.2 K. This dip in $R$ persists up to 8 K, indicating that superconductivity in the device is fully suppressed only above 8 K. Additionally, peaks in $R$ are present at high bias which we attribute to MAR \cite{KLAPWIJK19821657}. The first three peaks are indicated with $N= 1, 2, 3$ in the figure. That these features appear as peaks in $R$ is commonly observed in InAs JJ with epitaxial Al and indicates high interface transparency \cite{Kjaergaard2017}. The position of the MAR peaks depends strongly on temperature, as it is determined by the induced superconducting gap. Both the shifting of the MAR peaks as well as $R$ around zero bias current is qualitatively similar to devices with only Al as a superconductor, but the temperature range at which these processes occur is much larger \cite{Cheah2023}.

The size of the induced superconducting gap is determined from the voltage at which the MAR features occur. In Fig. \ref{f6}(f) $R$ measured at 600 mK is plotted as a function of source-drain voltage ($V_{\rm SD}$), which is obtained by integrating $R$ over $I_{\rm SD}$. In the inset of the figure the $V_{\rm SD}$ at which the first six MAR peaks occur are plotted as a function of 1/$N_{\rm MAR}$. This is fitted with the relation 
\begin{equation}
\label{eq:gap}
    eV=2 \Delta^* / N_{\mathrm{MAR}}
\end{equation}
Using \autoref{eq:gap} we find $\Delta^* =0.96$ meV, which is approximately five times larger than the induced gap typically found in Al based devices \cite{Cheah2023,Kjaergaard2017,Lee2019}. In Fig. \ref{f6}(g), the temperature shift of the most pronounced MAR peak ($N=2$) is investigated. The peak location is plotted as a function of temperature and fitted with the BCS relation \cite{Bardeen1957}
\begin{equation}
\Delta^*(T)=\Delta^*(0) \tanh \left(1.74 \sqrt{\frac{T_{\mathrm{c}}}{T}}-1\right)
\label{BCSstar}
\end{equation}
using $T_{\rm c}=7.8$ K obtained from measurements of a superconducting stripe on the same chip and fitting $\Delta^*$. The result of this fit was 0.99 meV for $\Delta^*$ at $T=0$ K, which agrees well with the previously found value for $\Delta^*$ of 0.96 meV at $T=0.6$ K.

Overall, the junction shows MAR features consistent with $\Delta^*$ of approximately 1 meV. In Appendix \ref{NbJJ} results are shown for a JJ with a width of 1 $\mu$m and a length of $100$ nm where Nb is the superconductor directly on the Al for which an even larger gap of around 1.6 meV was found. 

A similar experiment based on ex-situ deposition of NbTi on Al was previously performed by Drachman et. al \cite{Drachmann2017a}. In this study a value of 0.5 meV was found for $\Delta^*$, most likely due to damage induced into the Al film during native oxide removal. This difference highlights the importance of the in-situ deposition approach for formation of high transparent interfaces in hybrid structures. We increase the size of $\Delta^*$ and widen the parameter space which allows for detection of Andreev state related phenomena. Direct investigation of the density of states around $\Delta^*$, which would allow the detection of sub-gap states, is beyond the scope of this work.  

\section{Conclusions}
In summary, we have developed a semiconductor-superconductor hybrid heterostructure based on epitaxially stacked Nb and Al superconducting layers with a high critical magnetic field and temperature. We have optimized this material stack in terms of Al thickness and GaAs capping layers to form the high-quality interfaces between the superconductors and the semiconductor structure. These high-quality interfaces are achieved by in-situ depositing a large gap superconductor on top of the epitaxial Al layer. We have shown that this approach works for NbTi as well as Nb. The two superconducting layers have a clear epitaxial relationship and we have identified and analyzed two different grain combinations. 

The transport properties of a JJ fabricated on the optimised material stack with NbTi suggest high transparency at the semiconductor-superconductor interface. This is expected to be enabled by the epitaxial relationship between the materials. We have found that the device shows basic functionality such as gateability of the supercurrent and Fraunhofer interference pattern in a perpendicular magnetic field. Our detailed MAR analysis reveals strong interaction between the NbTi and the 2DEG mediated by the Al layer. Temperature-resolved measurements clearly show the influence of the NbTi, with signs of superconductivity in the device persistent up to 8 Kelvin. The induced gap was determined from MAR measurements to be approximately 1 meV, which represents a two-fold improvement over previous results \cite{Drachmann2017a}. The MAR peaks follow the BCS relation consistent with this induced gap. Similar results are obtained for devices based on Nb on Al. 

Our approach of in-situ depositing multiple superconducting layers and carefully optimizing the obtained interfaces could be extended towards other superconducting materials. A logical next step might involve even larger gap superconductors such as NbTiN. This path could lead to superconducting hybrid devices operating at liquid helium temperatures and could open possibilities for experiments requiring a large operating range in magnetic field. Additionally, our results could help in understanding the role of interface quality and the inverse proximity effect in other bilayer structures involving superconductivity \cite{maiani2024percolative,bakkers2022} or in the context of superconducting gap engineering \cite{Mcwen2024,Diamond2022,Marchegiani2022}. 

  \section*{Acknowledgements}
We thank Stefan Fält for technical support and M. Sousa (IBM) via the Binning and Rohrer Nanotechnology Center (BRNC) for TEM support. The authors also acknowledge ScopeM for their support and assistance. \textbf{Funding:} This project was funded  by the Swiss National Science Foundation (SNSF) and by the Swiss National Center of Competence in Research Quantum Science and Technology, QSIT. The authors also acknowledge the support of MEYS CR grant LM2018110 and the Czech Science Foundation grant No. 22-22000M. F.N. acknowledges support from the European Research Council (grant number 804273) and the Swiss National Science Foundation (grant number 200021201082).
\section*{Conflict of Interest}
The authors declare no conflict of interest.
\newpage
\appendix

\section{XRD results}

\begin{figure}[]
 
    \centering
    \includegraphics[width = 1\linewidth]{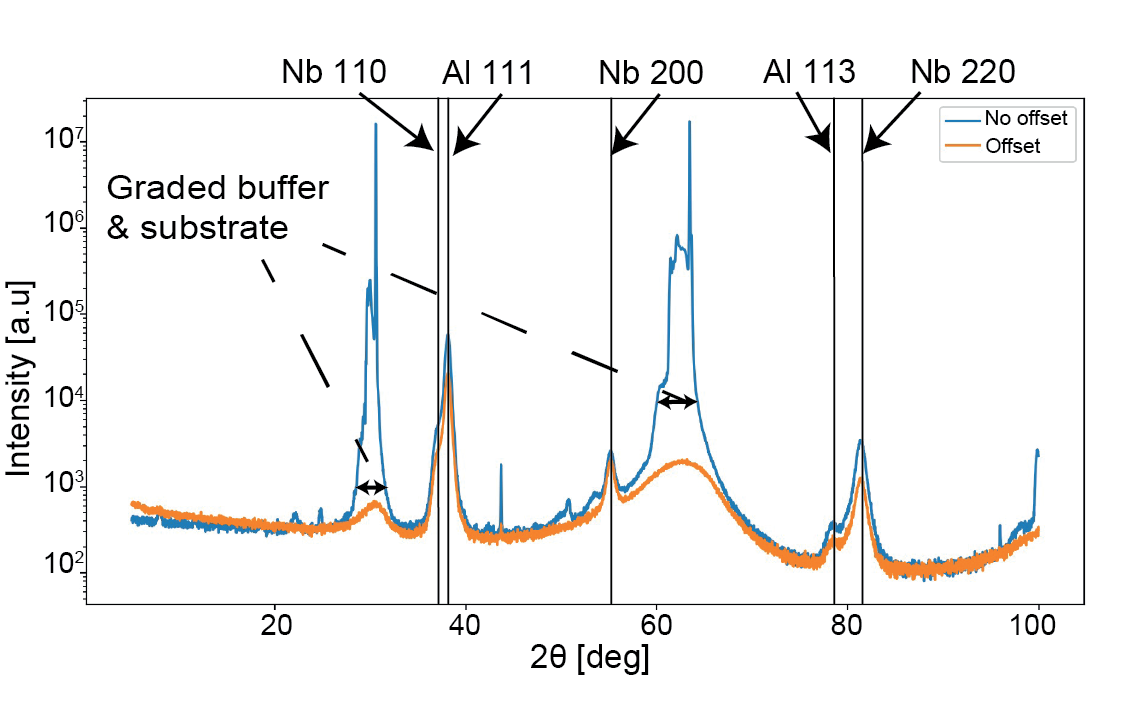}
    
    \caption{\textbf{XRD traces showing peaks associated with the various grain combinations identified in Fig. \ref{f1}.}}
    \label{xrd}
\end{figure}

XRD analysis of the InAs-Al-Nb based material is shown in Fig. \ref{xrd}. In order to differentiate between the monocrystalline substrate and the superconducting layers two traces are shown: one along the [001] axis and with a small offset with a slight offset with respect to this direction. The peaks that are associated with the monocrystalline semiconductor substrate as well as MBE-grown graded buffer structure are significantly less pronounced in the trace with the offset and are indicated in the figure. The peaks associated with the grain orientations described in Fig. \ref{f1} are labeled in the figure as well. We relate the observed Nb 110 and 220 peaks to the Nb [110] projection observed in the STEM. The Nb [100] projection is related to the observed 200 peak in the XRD data. Similarly, the observed Al 111 and 113 XRD peaks can be related to the [112] projection identified in STEM. We do not directly observe any peaks related to the Al [110] projection because the related XRD peaks overlap with the major substrate peak around 65$^\circ$. Overall, we conclude that the XRD data is consistent with the STEM observations and confirms the epitaxial relationship described in section \ref{epitaxy}.

\section{In migration}
\label{A2:indium}
\begin{figure}[]
 
    \centering
    \includegraphics[width = 1\linewidth]{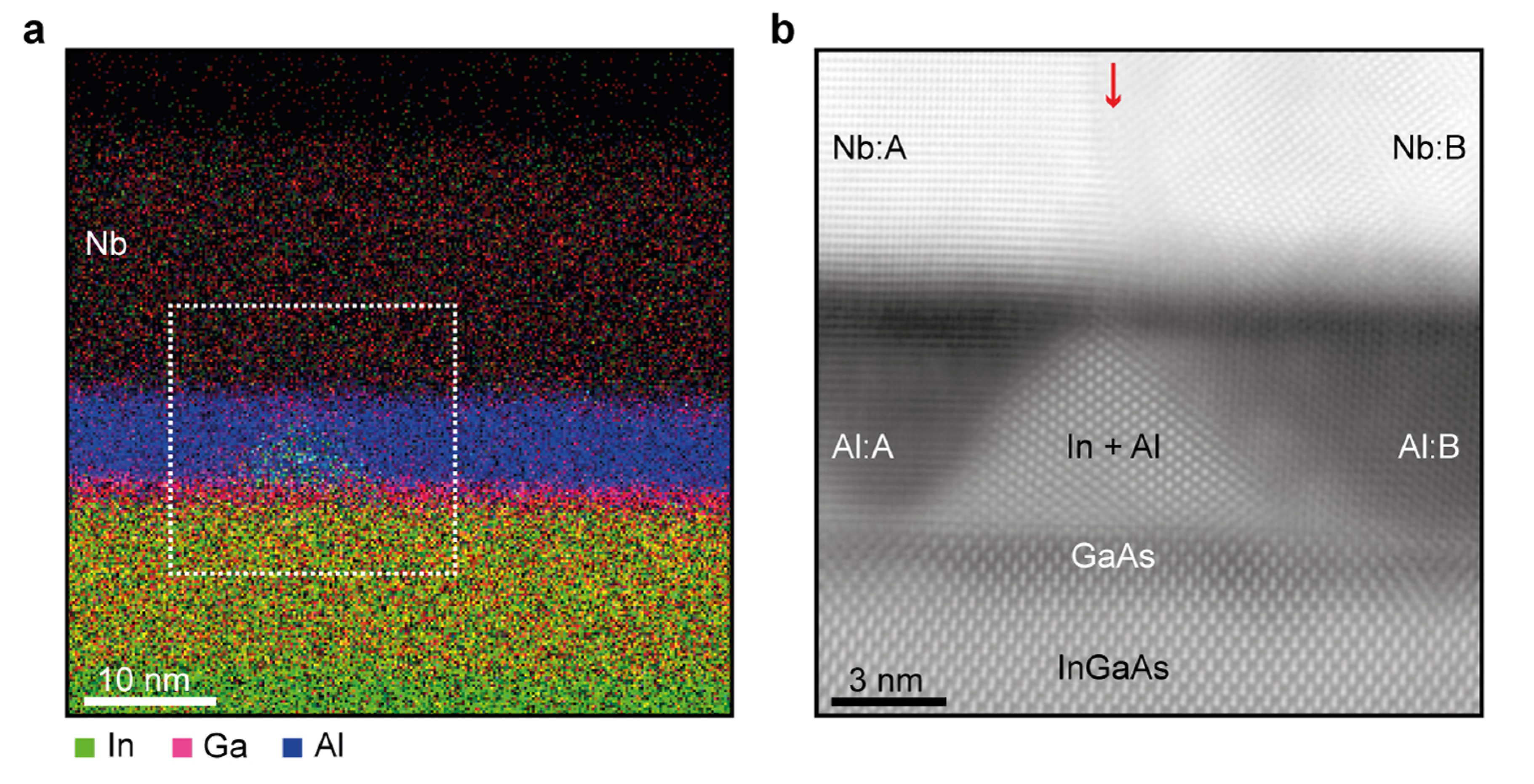}
    
    \caption{\textbf{HAADF STEM EELS images of the observed triangular indium migration that occurs if the semiconductor heterostructure is not adequately capped with GaAs capping layers.  } \textbf{a} An elemental map obtained using an in-situ EELS detector that shows the migration of In trough the Al layer. \textbf{b} HAADF STEM close up image of the In triangle. It can be clearly seen that the In interruption causes the Al grain orientation to switch (going from grain combination A on the left to grain combination B on the right). The Nb follows this switch in grain orientation, further confirming the one-to-one correspondence and epitaxial relationship between the two superconductors.}
    \label{In_migration}
\end{figure}

 Figure \ref{In_migration}(b) shows a HR BF STEM image of a triangular inclusion and in Fig. \ref{In_migration}(a) an EDX elemental map of that same region shows the associated increased In content in the Al layer. The two observed grain orientations are also indicated in Fig. \ref{In_migration}(b) and it can be clearly seen that the In rich region introduces a grain boundary in the superconducting layers (going from grain combination A on the left to grain combination B on the right). This also results in a grain boundary in the Nb superconducting layer, which further confirms its epitaxial relationship to the Al layer. 

\section{JJ fabrication}
\label{JJfab}
To fabricate the Josephson junctions in this study, we mostly follow the standard fabrication procedure reported in Ref.~\cite{Cheah2023}. Before the Al wet etch in Transene-D we remove the Nb/NbTi layer via reactive ion etching using SF$_6$ as etchant.

\section{Transport measurements on Nb-Al based JJ}
\label{NbJJ}
\begin{figure}[]
 
    \centering
    \includegraphics[width = 1\linewidth]{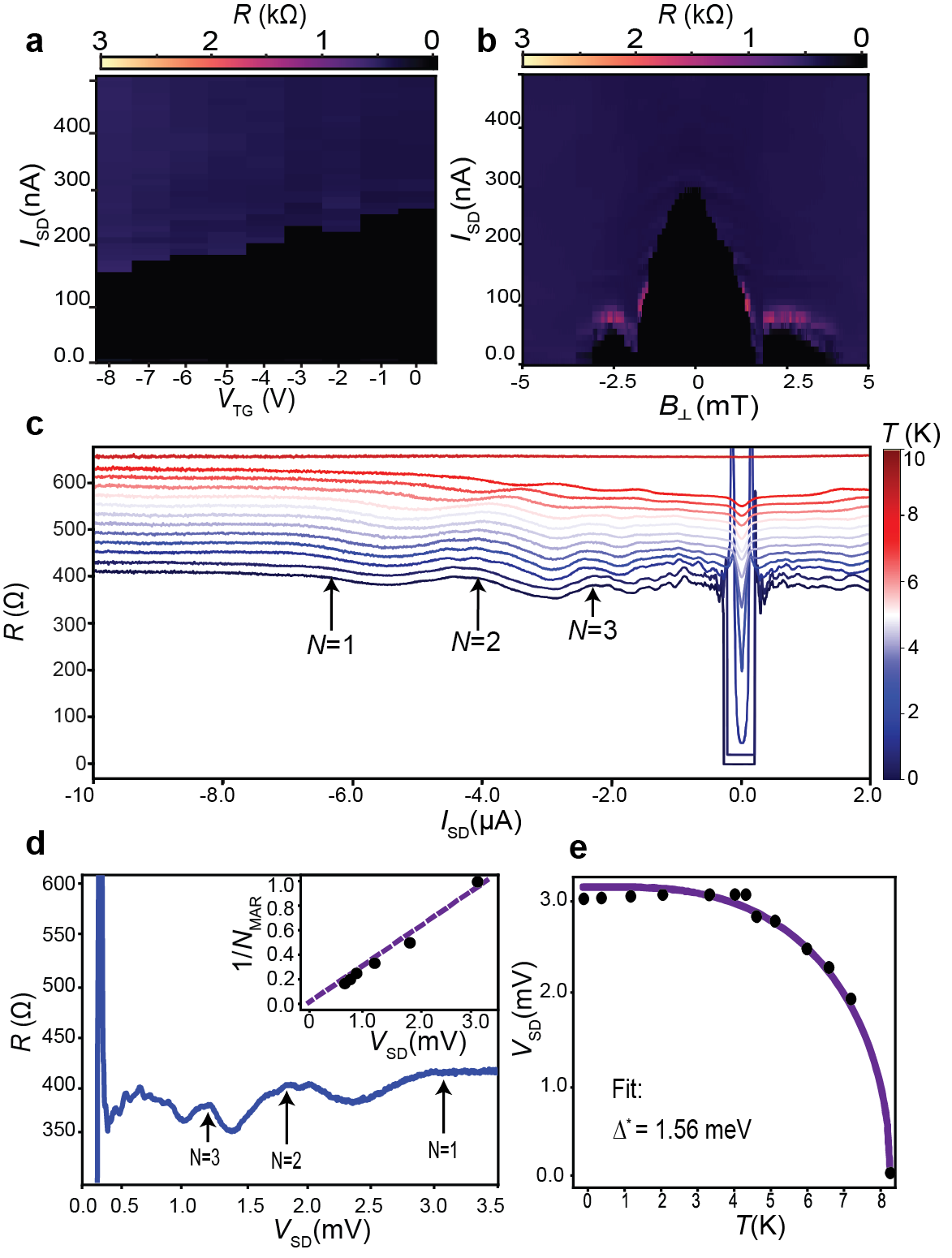}
    
    \caption{\textbf{Transport measurements on a JJ fabricated from a wafer with 50 nm Nb directly on the epitaxial Al.} \textbf{a} the dependence of the resistance on the bias current and gate voltage.  In \textbf{b} the dependence of the resistance R on the bias current and applied magnetic field is shown. \textbf{c} The resistance of the JJ as a function of DC bias voltage for increasing temperature. Indicated are the peaks that we assign to MAR.  \textbf{d} The resistance at 1300 mK as a function of $V_{\rm SD}$ with in the figure inset the first six MAR peaks plotted as a function of 1/$N_{\rm MAR}$. The peak locations are fitted with an induced gap of 1.64 meV. \textbf{e} The temperature related shift of the N=1 MAR peak and fit with equation \ref{BCSstar}.}
    \label{fig:NbJJ}
\end{figure}

Figure \ref{fig:NbJJ} shows transport measurements of a JJ made from the heterostructure material with 50 nm Nb deposited in-situ directly on the epitaxial Al layer. The junction has the same dimensions as the one presented in Fig. \ref{f6} and the same measurement techniques are used. All measurements are done in a dilution refrigerator with a base temperature of 10 mK. 

Overall, very similar data is obtained compared to the device presented in Fig. \ref{f6}. One difference that is shown in Fig. \ref{fig:NbJJ}(a) is that this device only reacted very weakly to the applied gate voltage and therefore complete pinch off of the super current could not be reached. This is most likely a fabrication related problem which is specific to this device. The obtained Fraunhofer pattern in Fig. \ref{fig:NbJJ}(b) is very similar to the device in Fig. \ref{f6}, but with a factor of two difference in periodicity which could relate to fabrication related variations in JJ length. Additionally, a difference in penetration depth of the NbTi compared to the Nb could play a role as it will result in decay of the order parameter over a longer length scale, which affects the junction length. However, this effect is difficult to quantify since the penetration depth values for these thin films will differ from known bulk values. The observed MAR features again persist up to high bias current and the location of the peaks can be fitted with \autoref{eq:gap} finding an induced gap energy of 1.64 meV. Similarly, the $N=1$ feature can be tracked as a function of temperature and fitted with equation \autoref{BCSstar} to find an induced gap of 1.56 meV. We interpret these results as a confirmation that our approach of stacking superconductors and optimising the interfaces is effective for multiple superconductors and multiple devices. 

\section{2DEG characterisation}
\begin{figure}
 
    \centering
    \includegraphics[width = 1\linewidth]{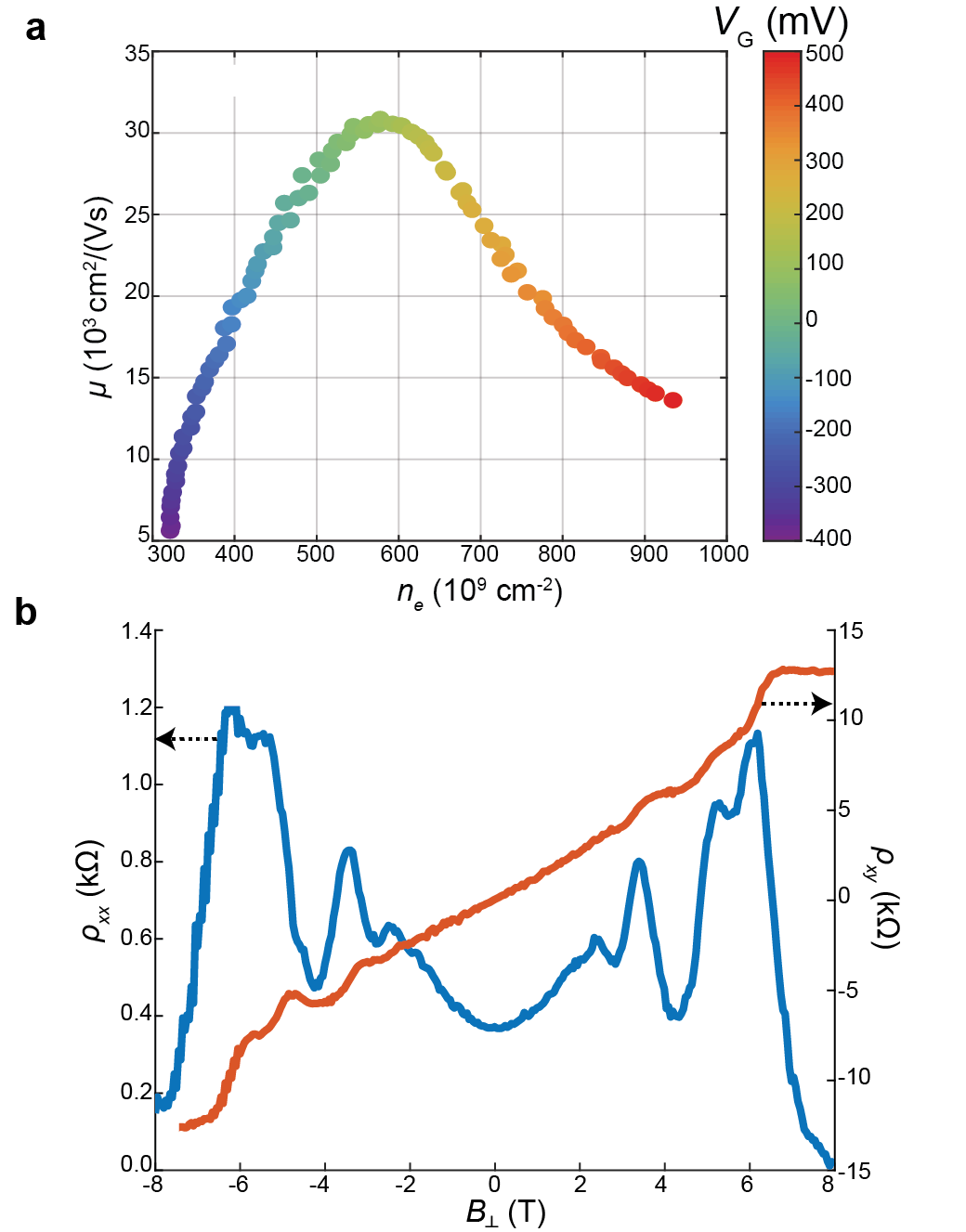}
    
    \caption{\textbf{Low temperature DC transport characterization of a Hall bar showing the mobility and density of the 2DEG as well as the Shubnikov-de Haas oscillations and Hall plateaus.  }}
    \label{2DEG}
\end{figure}

Magneto-transport characterisation was performed on Hall bars fabricated from the optimised semiconductor heterostructure with 2 ML of GaAs capping, 7 nm of Al and 60 nm of Nb on top to evaluate the mobility and carrier density in the 2DEG. Both superconductors were etched away and a standard Hall bar was fabricated with a top gate. Measurements shown in Fig. \ref{2DEG} are done at 1.4 K and in Fig.\ref{2DEG}b the top gate voltage was set to zero. The maximum mobility is similar to what is typically achieved in state of the art 2DEGs with epitaxial Al alone \cite{Coraiola2023,Lee2019}, indicating that the additional Nb deposition has limited negative effect on the 2DEG quality. 

\section{Superconductor DC transport characterization}
Low temperature DC transport measurements were performed in the Van der Pauw geometry using standard lockin techniques. By controlling the temperature and magnetic field, the critical temperature $T_{\rm c}$ as well as critical out-of-plane magnetic field $B_{\rm c2}$ are extracted. Figure \ref{NbTi}(c) shows the results for a semiconductor-superconductor hybrid structure with 7 nm of Al and 60 nm of Nb as well as a reference structure with the same semiconductor heterostructure but with 60 nm Nb deposited direction on the semiconductor. Both films far exceed the known critical temperature and out-of-plane magnetic field of state of the art epitaxial Al films of 1.5 K and 164 mT respectively \cite{Mayer2019}. Additionally, two more films are shown (with and without the 7 nm of Al) where the top superconductor is NbTi rather than pure Nb. 

Firstly, it can be seen that both the Nb and NbTi film without an Al interlayer have a $T_{\rm c}$ of around 9.3 K, which is very close to the maximum achievable bulk value \cite{Roberts1976}. This is a clear sign of a low impurity concentration in our superconducting Nb film, which is known to reduce the $T_{\rm c}$. If we now compare this $T_{\rm c}$ of 9.3 K to the material stack with the 7 nm Al layer included, we observe a reduction of about 1.9 K to 7.4 K, which we attribute to the inverse proximity effect introduced by the Al. This effect, which can be thought of as part of the Cooper pair wave function leaking into the Al, has been observed in other experiments, for example, in Nb-Ni bilayers \cite{Skryabina2019} but also Ti-Au bilayers \cite{Nagayoshi2022}. The NbTi films are not strongly affected by the presence of the Al layer, which can be attributed to the significantly shorter coherence length. 

Additionally, the values of $B_{\rm c2}(T)$ at low temperature clearly indicate that the Al interlayer increases the critical field significantly. While the reference film of Nb as a critical field of approximately 2 T at 20 mK, the material stack with 7 nm Al interlayer already has a critical magnetic field of 2.5 T at 2 K. We attribute this increase to the pinning of magnetic vortices at the Al-Nb interface consequently increasing the critical magnetic field.

\bibliographystyle{ieeetr}
\bibliography{papers}
\end{document}